\documentclass[doublecol]{epl2}

\usepackage{amsmath}
\usepackage{amstext}
\usepackage{amssymb}
\usepackage{latexsym}

\newcommand{\beq}{\begin{equation}}
\newcommand{\eeq}{\end{equation}}
\newcommand{\ket}[1]{\left\vert#1\right\rangle}
\newcommand{\bra}[1]{\left\langle#1\right\vert}
\newcommand{\Ham}{\mathcal H}
\newcommand{\ave}[1]{\left\langle#1\right\rangle}

\title{Negative differential conductivity in far-from-equilibrium \\
quantum spin chains}

\author{Giuliano Benenti\inst{1,2} \and Giulio Casati\inst{1,2,3} \and Toma\v z Prosen\inst{4} \and Davide Rossini\inst{5}}

\institute{
\inst{1} CNISM, CNR-INFM, and Center for Nonlinear and Complex Systems,
         Universit\`a degli Studi dell'Insubria,\\
         Via Valleggio 11, 22100 Como, Italy

\inst{2} Istituto Nazionale di Fisica Nucleare, Sezione di Milano,
         Via Celoria 16, 20133 Milano, Italy

\inst{3} Centre for Quantum Technologies,
         National University of Singapore, Singapore 117543

\inst{4} Physics Department, Faculty of Mathematics and Physics,
         University of Ljubljana, Ljubljana, Slovenia

\inst{5} International School for Advanced Studies (SISSA),
         Via Beirut 2-4, I-34014 Trieste, Italy}


\abstract{
We show that, when a finite anisotropic Heisenberg 
spin-1/2 chain in the gapped regime is driven far from equilibrium,
oppositely polarized ferromagnetic domains build up at the edges 
of the chain, thus suppressing quantum spin transport. 
As a consequence, a negative differential conductivity regime arises, 
where increasing the driving decreases the current. 
The above results are explained in terms of magnon localization 
and are shown to be structurally stable against breaking of integrability.
}

\begin{document}

\maketitle

Quantum transport properties of low-dimensional materials are currently
an object of intensive theoretical and experimental
research~\cite{zotos:BOOK,wolf01,spin:RMP}.
Low-dimensional systems are interesting for theoretical investigations,
as they admit ordering tendencies, leading to collective quantum states
that are difficult to realize in three-dimensional systems.
Experimentally, unconventional transport properties have been 
reported, including unusually high thermal conductivity in quasi-one 
dimensional magnetic compounds and ballistic spin transport in 
magnetic chains (see Ref.~\cite{zotos:BOOK} and references therein). 
Understanding the transport properties of such low-dimensional
strongly correlated systems is a challenging open problem.
So far, most of the theoretical studies concentrated on the
close-to-equilibrium situation by using the linear-response
formalism~\cite{meier03,zotos99,prelovsek04}, while almost nothing is known
about the physics of such systems far from equilibrium.
On the other hand, new quantum phases and interesting physical phenomena
may appear in the far-from-equilibrium regime~\cite{tomaz}.

In order to drive an interacting quantum system far from equilibrium,
it is necessary to strongly couple it to some macroscopic reservoirs.
Theoretical description of such situation usually relies on
a master equation approach for the density matrix of the system,
where the dissipative term depends on the coupling of the model to
the reservoirs.
An analytical treatment of such situations in non-trivial many-body systems
is generally unfeasible~\cite{njp08}, while numerical simulations
are highly demanding and, as far as we know, only very few examples
are discussed in the literature~\cite{saito}.

In this paper we propose a conceptually simple model of an anisotropic
one-dimensional quantum Heisenberg spin-$1/2$ chain coupled to a pair of 
magnetization reservoirs.
We consider a spin chain driven far from equilibrium, much beyond the
linear-response regime: we show that a {\it long-range spin ordering
into ferromagnetic domains} is induced, independently of the ferromagnetic
or antiferromagnetic nature of the spin-spin coupling. This cooperative,
many-body quantum state hampers spin flips, thus strongly suppressing the current.
Therefore, the spin current, which at small driving obeys Ohm's law,
for sufficiently strong driving exhibits {\it negative differential
conductivity} (NDC), namely increasing the driving decreases the current.

A NDC effect has been already predicted and observed in nanoscopic objects,
such as single molecules, short nanotubes, or quantum dots weakly coupled
to metallic electrodes (see, e.g., Refs.~\cite{thielmann05,elste06}
and references therein). Transport properties in such systems are usually
studied by considering effective models of a few single-particle levels.
Conversely, we remark that in our case the NDC has a different
origin and unveils a new phenomenon: it arises as an outcome of a beautiful
interplay between coherent many-body quantum dynamics of the spin chain
and incoherent spin pumping.

Our system is constituted by $N$ interacting quantum spins $1/2$, whose
dynamics is described by an XXZ Heisenberg exchange Hamiltonian:
\beq
   \Ham_S = \sum_{k=1}^{N-1} \Big[ J_x ( \sigma^x_k \sigma^x_{k+1} +
   \sigma^y_k \sigma^y_{k+1} ) + \ J_z \sigma^z_k \sigma^z_{k+1} \Big] \, ;
   \label{eq:Heisenberg}
\eeq
here $\sigma^\alpha_k$ ($\alpha = x,y,z$) are the Pauli matrices of
the $k$-th spin, and $\Delta \equiv J_z / J_x$ denotes the $xz$ anisotropy.
Both ends of the spin chain are coupled to magnetic baths, that is,
to magnets acting as reservoirs for the magnetization~\cite{meier03}.
Within the Markovian approximation, which holds provided the
bath relaxation time scales are much shorter than the time scales
of interest for the system's dynamics, the time evolution of the 
system's state density matrix $\rho(t)$ follows the Lindblad
master equation~\cite{breuer}:
\beq
   \frac{\partial \rho}{\partial t} = -\frac{i}{\hbar} [ \Ham_S, \rho]
   - \frac{1}{2} \sum_{m=1}^4 \{ L_m^\dagger L_m , \rho \}
   + \sum_{m=1}^4 L_m \rho L^\dagger_m \, ,
   \label{eq:MasterEq}
\eeq
where 
\beq
\begin{array}{ll}
   L_1 = \sqrt{\Gamma \mu_L}          \, \sigma^+_1 \,, &
       L_2 = \sqrt{\Gamma (1-\mu_L)}  \, \sigma^-_1, \\\\
   L_3 = \sqrt{\Gamma \mu_R}          \, \sigma^+_N \,, &
       L_4 = \sqrt{\Gamma (1-\mu_R)}  \, \sigma^-_N
\end{array}
   \label{eq:Lindblad}
\eeq
are four Lindblad operators, 
$\sigma^\pm_k=(\sigma^x_k\pm i \sigma^y_k)/2$ are the raising and 
lowering operators, while $[\cdot,\cdot]$ and $\{ \cdot, \cdot \}$
denote the commutator and the anti-commutator, respectively.
Hereafter we use dimensionless units, by setting $\hbar=1$, $J_x=1$.
The Lindblad operators $L_1, \, L_2$ act on the leftmost spin of the chain
($k=1$), while $L_3, \, L_4$ on the rightmost one ($k=N$).
The dimensionless parameters $\Gamma$ and $\mu_L$ ($\mu_R$) play the role
of the system-reservoir coupling strength and of a left (right) chemical
potential, respectively. 
In other words, $2\mu_{L,R}-1 \in[-1,1]$ is
the corresponding bath's magnetization per spin in dimensionless units.

Given the conceptual simplicity of our model, various experimental 
implementations could be envisaged. For instance, one could consider
molecular spin wires~\cite{bogani08,ghirri07} with each boundary
coupled to a magnetic impurity, where the desired populations
$\{\mu_{L,R},1-\mu_{L,R}\}$ of up/down spins can be
pumped by means of electromagnetic fields.
We choose a symmetric driving: $\mu_{L,R} = \frac{1}{2} (1 \mp \mu)$, so that
$\mu \equiv \mu_R - \mu_L \in [0,1]$ is a single 
parameter controlling the driving strength.
When $\mu$ is small we are in the linear-response regime, while in the limiting
case $\mu = 1$ (corresponding to $\mu_L = 0, \mu_R = 1$) the left (right)
bath only induces up-down (down-up) spin flips.
This last case can be thought of as two spin-polarized chains coupled to the
system. For instance, one could have two oppositely polarized ferromagnets
for the left and the right baths.

The time evolution of the master equation~\eqref{eq:MasterEq}
has been numerically simulated by employing a Monte Carlo wave function
approach, based on the technique of {\it quantum 
trajectories}~\cite{dalibard92,gabriel}.
We assume that there exists a single out-of-equilibrium steady state
and study the stationary spin current $\langle j \rangle$, 
which is calculated by looking at the left
bath and summing up all up-down flips minus all down-up flips, and then
dividing by the simulation time. Due to the conservation of the total
magnetization in the Hamiltonian model~\eqref{eq:Heisenberg},
after the convergence time
this current is precisely equal to the analogous quantity computed at the
right bath and also equals the expectation value
of $j_k = J_x ( \sigma^x_k \sigma^y_{k+1} - \sigma^y_k \sigma^x_{k+1} )$
for any $k=1,\ldots,N-1$.

\begin{figure}[!t]
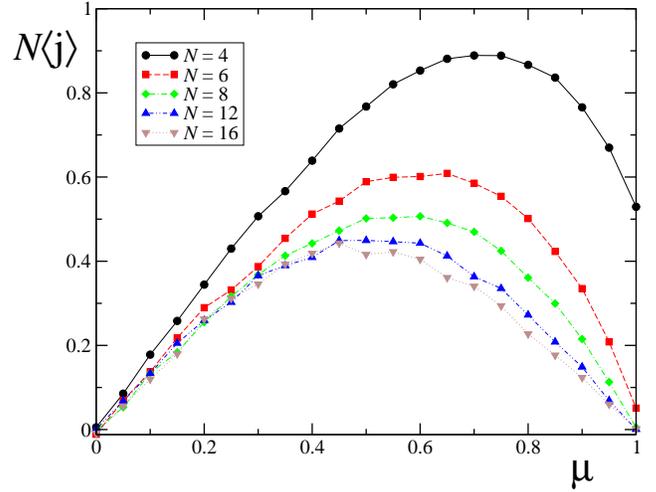

    \onefigure[scale=0.35]{CurrL}
    \caption{(Colour on-line). Stationary spin current $\langle j \rangle$
      as a function of the driving strength $\mu$, with anisotropy $\Delta = 2$
      and system-baths coupling strength $\Gamma = 4$.
      We rescaled the $y$-axis with the system size $N$ in such a way that
      the small-$\mu$ Ohmic behaviour can be clearly appreciated.}
    \label{fig:CurrL}
\end{figure}

Quite surprisingly, for an anisotropy $|\Delta|>1$, the
stationary spin current exhibits a negative differential conductivity 
phenomenon, as shown in Fig.~\ref{fig:CurrL}.
Namely, in the linear-response regime of small $\mu$ we find an Ohmic
behaviour $\langle j \rangle \propto \mu /N$
of the spin current $\langle j \rangle$; the current reaches
a maximum at a certain value $\mu^\star$, and, further increasing $\mu$,
it decreases, thus exhibiting a negative differential.
At maximum driving strength $\mu=1$, the current drops {\em exponentially} with $N$.
While the small-$\mu$ Ohmic behaviour is consistent with recent
linear-response numerical results~\cite{prelovsek04},
the NDC effect is completely unexpected.

\begin{figure}[!t]
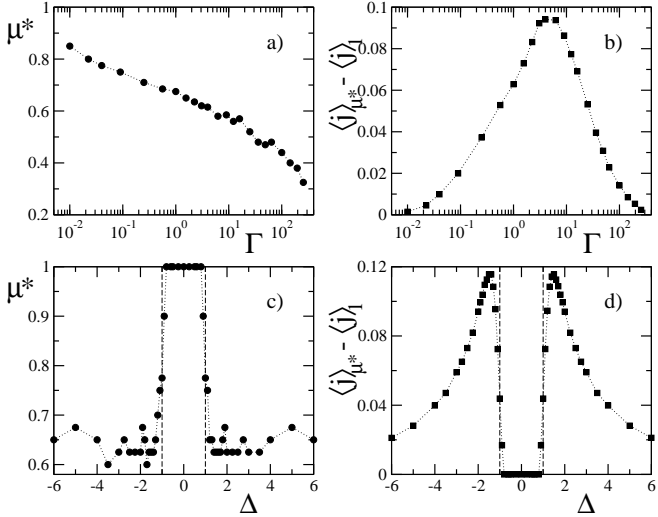

    \onefigure[scale=0.32]{Curr_Gvar_bn}
    \caption{Driving strength $\mu^*$ at which the spin current exhibits
      a maximum (left) and current at $\mu^*$ minus current at $\mu=1$ (right),
      for a chain of $N=6$ spins.
      Top panels are for $\Delta=2$, bottom panels for $\Gamma = 4$.}
    \label{fig:Curr_Gvar}
\end{figure}

NDC is robust upon the variation of the number $N$ of spins
(Fig.~\ref{fig:CurrL}) and of the system parameters $\Gamma$ and $\Delta$
(Fig.~\ref{fig:Curr_Gvar})~\footnote{
  Furthermore, the phenomenon has been hinted by one of us in a different
  context of kicked open quantum dynamics:
  Prosen T., unpublished notes available at
  {\tt arXiv:0704.2252 [quant-ph]}.
}.
By increasing $\Gamma$ over more than four orders of magnitude, 
from $\Gamma = 10^{-2}$ to $\Gamma = 3 \times 10^2$, we find 
[see Fig.~\ref{fig:Curr_Gvar}a)] that the driving strength
$\mu^\star$ at which the current reaches its maximum shifts from
$\mu^\star\approx 0.9$ to $\mu^\star\approx 0.3$, thus considerably
reducing the range of validity of the linear-response regime.
The maximal current drop, measured by 
$\langle j \rangle_{\mu=\mu^\star} - \langle j \rangle_{\mu=1}$,
is obtained for $\Gamma\approx 5$ [see Fig.~\ref{fig:Curr_Gvar}b)].
When $\Gamma\ll 1$ the ``contact resistance'' 
$1/\Gamma$ becomes large and the current is small for any value of $\mu$.
On the other hand, for $\Gamma\gg 1$ we are in 
the quantum Zeno regime~\cite{pascazio},
where the coupling to the reservoirs is strong and freezes the system's
dynamics, thus drastically reducing the spin current. 

In the bottom plots of Fig.~\ref{fig:Curr_Gvar},
the driving strength $\mu^\star$ and the current drop 
$\langle j \rangle_{\mu^\star} -\langle j \rangle_1$
are shown as a function of the anisotropy $\Delta$.
For $|\Delta| < 1$ the negative differential conductivity effect is absent and
the linear-response regime $\langle j \rangle\propto \mu$ can be extended
up to $\mu=1$, with an ideally conducting behaviour,
so that the spin current is independent of the system size~\cite{zotos99}.
On the other hand, NDC is observed at any $|\Delta| > 1$
[see Fig.~\ref{fig:Curr_Gvar}c)].
Since spin transport along the chain is suppressed when $|\Delta|\gg 1$,
there exists a value of the $xz$ anisotropy, $|\Delta|\approx 1.5$
[see Fig.~\ref{fig:Curr_Gvar}d)], at which the current drop is maximal.
From the above analysis we infer that, for a chain of 
$N=6$ spins, the optimal working point for the observation 
of the NDC phenomenon is $\Gamma\approx 5$, $|\Delta|\approx 1.5$. 

\begin{figure}[!t]
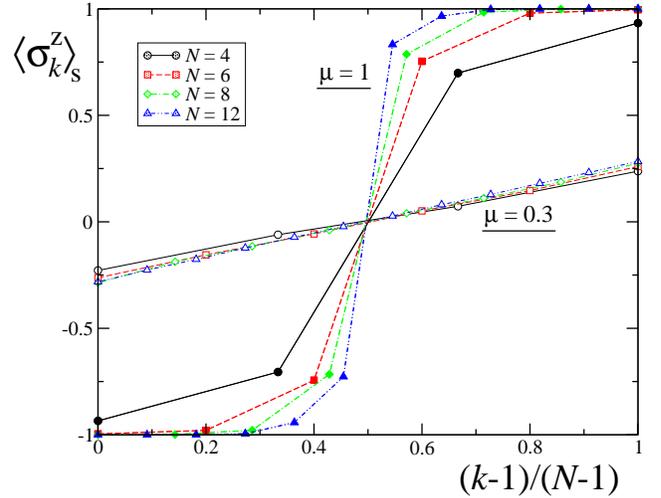

    \onefigure[scale=0.35]{SpinZ}
    \caption{(Colour on-line). Spin magnetization profiles
      $\langle \sigma^z_k \rangle_s$
      versus the scaled spin index coordinate $\frac{k-1}{N-1}$ at
      $\Delta = 2$, $\Gamma=4$, for driving strengths 
      $\mu=0.3$ (empty symbols) and $\mu=1$ (filled symbols).}
    \label{fig:SpinZ}
\end{figure}

Fig.~\ref{fig:SpinZ} displays the stationary spin magnetization  
profiles $\ave{\sigma^z_k}_s$ along the chain, for the parameter values
of Fig.~\ref{fig:CurrL}.
In the linear-response regime, we observe a constant linear gradient,
with the magnetizations $\langle \sigma^z_{1,N} \rangle_s$ 
of the two borderline spins close to the bath magnetizations
$\langle \sigma^z_{L,R} \rangle = 2 \mu_{L,R} -1=\mp \mu$.
This behaviour is typical of normal Ohmic conductors~\cite{lepri}.
Interestingly, in the limiting case $\mu=1$ we notice the appearance
of a stationary state characterized by two almost ferromagnetic domains,
that are polarized as the nearest reservoir and whose {\em relative}
width increases with the system size.
These ferromagnetic regions are responsible for strongly inhibiting
spin flips, and therefore for suppressing the spin current.
We remark that we found no differences between a ferromagnetic ($J_z<0$)
and an antiferromagnetic ($J_z>0$) spin coupling, and in particular we 
observed the same stationary spin profiles.
In the first case the formation of ferromagnetic domains may 
be somewhat intuitive, and indeed it can also be observed in the ground state 
of the autonomous XXZ chain exposed to two oppositely oriented static 
magnetic fields applied at the chain edges~\cite{alcaraz95,matsui96}.
On the other hand, the build up of ferromagnetic domains 
is is a priori not obvious for an antiferromagnetic coupling.
Indeed in this last case, for the autonomous model~\eqref{eq:Heisenberg},
ferromagnetic domains correspond to a highly excited state; 
from the point of view of the Hamiltonian system,
{\it the net effect of the baths 
is that of pumping energy into the system, while
leading to a stationary state with very low entropy}.

\begin{figure}[!t]
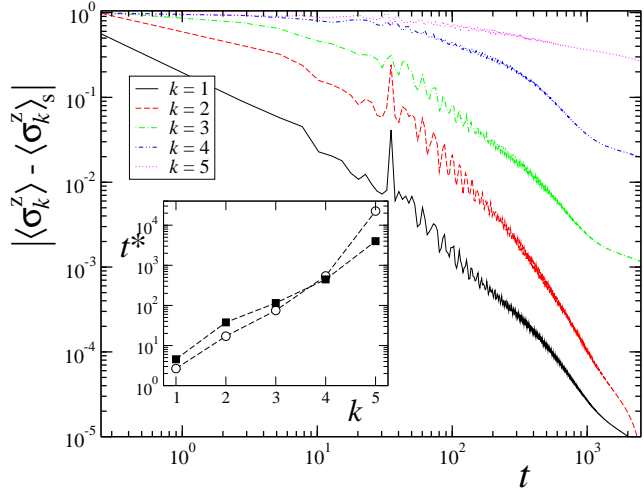

    \onefigure[scale=0.35]{Relax}
    \caption{(Colour on-line). Time approach to the equilibrium value of the spin
      magnetization of the leftmost spins,
      for $\mu=1$, $N=12$, $\Delta = 2$, $\Gamma=4$.
      From bottom to top: $k=1,2,...,5$.
      In the inset we plot the times $t^*$ at which the magnetization
      reaches a given fraction of the corresponding stationary state value:
      $\langle \sigma^z_k \rangle = \frac{9}{10} \langle \sigma^z_k \rangle_s$,
      as a function of the spin index $k$ (empty circles).
      The case with a single bath coupled to the first spin ($k=1)$
      is also shown (filled squares).}
    \label{fig:Relax}
\end{figure}

As shown in Fig.~\ref{fig:Relax}, the time scale needed to reach the stationary
state at $\mu=1$ is exponentially long.
This is the key observation on which our intuitive explanation of the NDC
phenomenon relies, as we shall qualitatively explain below.
Since at small $\mu$ the current grows linearly, it is sufficient to show that
the current is suppressed in the limiting case $\mu=1$ to conclude that, due to
the continuity of $\langle j \rangle_{\mu}$, a region of negative differential
conductivity exists. Therefore, we focus on the $\mu=1$ case. 

First, it is instructive to consider a situation in which the system
is coupled to a single, fully polarized reservoir, $\mu_L=0$.
Regardless of the anisotropy $\Delta$, the stationary state 
is pure and ferromagnetic, namely
$\ket{\downarrow \downarrow \cdots \downarrow}\bra{\downarrow \downarrow \cdots \downarrow}$,
since the Hamiltonian~\eqref{eq:Heisenberg} conserves the overall magnetization
while at the left boundary of the chain only the lowering operator
$L_2\propto \sigma_1^-$ acts. 
As Fig.~\ref{fig:Relax} (inset) shows, at $|\Delta|>1$ also for the single-bath
case the convergence of $\langle \sigma^z_k \rangle$ to the equilibrium value
$\langle \sigma^z_L \rangle =-1$ requires a time scale $t^\star$ which grows
{\em exponentially} with the spin-bath distance $k$
(this sharply contrasts with the ideally conducting case
$|\Delta|<1$, where $t^\star$ only grows {\em linearly} with $k$).
Moreover, if an up-polarized reservoir is added at the right boundary,
at $|\Delta|>1$ the relaxation times $t^\star$ for the spins closer to the
left than to the right boundary do not change significantly.
This implies that, in our two-baths model, the spin polarization
is practically affected only by the nearest bath. 

The above results can be explained in terms of localization of
one-magnon excitations. Given a ferromagnetic state
$\ket{0} \equiv \ket{\downarrow \downarrow \cdots \downarrow}$,
one-magnon excitations have the general form 
$\sum_{k=1}^N c_k \ket{k}$, where $\ket{k}=\sigma^+_k \ket{0}$
describes the state with the $k$-th spin flipped.
If the XXZ chain of Eq.~\eqref{eq:Heisenberg}
has open boundary conditions, there is an energy gap $2|J_z|$
between states $\ket{1}$ and $\ket{N}$ 
(spin-flip excitations at the boundaries) and states
$\ket{2}$, $\ket{3}$,..., $\ket{N-1}$.
Indeed one has $\bra{1} \Ham_S \ket{1} = \bra{N} \Ham_S \ket{N}
= (N-3) J_z$, and $\bra{2} \Ham_S \ket{2} = \ldots = \bra{N-1} \Ham_S \ket{N-1}
= (N-5) J_z$.
Only nearest-neighbour spin-flipped states are coupled, with a
coupling $\bra{k} \Ham_S \ket{k+1} = 2 J_x$,
therefore the Hamiltonian~\eqref{eq:Heisenberg} in the one-magnon
basis $\{ \ket{1}, \ket{2}, \ldots , \ket{N} \}$ is a tridiagonal
matrix that can be diagonalized exactly, in the limit of large $N$.
One finds that there always exist at least $N-2$ delocalized solutions.
If $\vert \Delta \vert =\vert J_z/J_x \vert>1$, 
both for ferromagnetic and antiferromagnetic
coupling, two peculiar eigenstates emerge: these ``molecular orbitals''
read $|\psi_{\pm}\rangle\approx \frac{1}{\sqrt{2}}
(|\psi_L\rangle\pm |\psi_R\rangle$), where the states 
$|\psi_{L,R}\rangle$ are centered at sites $1$ and $N$, respectively,
with a localization length $\ell\approx 1/ \ln \vert \Delta \vert$.
The gap between the two corresponding energy levels shrinks exponentially
with the system size, so that the coherent tunneling between the two
border sites requires a time scale that increases exponentially with $N$.
In practice, this means that spin-flip excitations created at the border
of the chain remain exponentially localized when $\vert\Delta\vert > 1$,
over a localization length $\ell$.

Consider now the action of the two baths. 
As can be clearly seen from Fig.~\ref{fig:Relax}, intermediate states
in the relaxation to the stationary spin profile have two opposite-oriented
ferromagnetic domains close to the baths.
In order to enlarge them, spin-flip excitations should be
propagated, through $\sigma^x_k\sigma^x_{k+1}$ and $\sigma^y_k\sigma^y_{k+1}$ 
exchange couplings of Hamiltonian~\eqref{eq:Heisenberg},
across the ferromagnetic domains to the chain boundaries.
Suppose, for instance, that we have the leftmost $m$ spins down and the $(m+1)$-th
spin up and that this excitation propagates to the left bath; then the bath
can flip this spin down, thus ending up with a ferromagnetic domain with
$m+1$ leftmost spins down.
The crucial point is that the one-magnon propagation is exponentially
localized at $|\Delta|>1$. This explains why (i) equilibration needs
exponentially long time scales and (ii) only the nearest bath is felt.
We stress, however, that negative differential conductivity is observed after
quite short time scales ($t\sim 10^2$), independently of the chain length.
Indeed, it is sufficient to create a very small ferromagnetic
region close to a bath, to strongly suppress current.
For example, this {\em spin-blockade} mechanism can be clearly
seen at $\mu \sim 0.9-0.95$, where only a couple of outer spins
reach magnetization values close to $\pm 1$, but still the current is far below
the peak value at $\mu^\star$.

We finally mention that the presence of NDC is not related
to integrability of the Heisenberg model.
Indeed, we performed numerical simulations on XXZ
chains in presence of a staggered magnetic field along the $z$ direction,
$B_{\rm stag}(k)=(-1)^k B$.
This system exhibits a transition from integrability 
to quantum chaos~\cite{haake}, when increasing the field strength $B$.
We found that the NDC phenomenon is insensitive to such a transition.

We point out that our model might be relevant 
for far-from-equilibrium electronic transport through
molecular wires that are embedded between two electrodes~\cite{joachim00}.
This is suggested by the fact that the XXZ chain can be mapped by the
Jordan-Wigner transformation~\cite{mattis} into a tight-binding $t-V$ model of 
spinless fermions moving with a hopping amplitude $t=2 J_x$ 
on a $N$-site one-dimensional lattice,
with nearest neighbour interaction of strength $V=4 J_z$.
Here the spin current of the XXZ model becomes a charge current,
the Lindblad operators~\eqref{eq:Lindblad} are mapped 
into operators injecting/extracting electrons at the lattice boundaries,
and $\frac{1}{2} (1+ \langle \sigma_k^z \rangle)$ into
the charge density at site $k$. In particular, the ferromagnetic domains of
Fig.~\ref{fig:SpinZ} imply a {\em phase separation} in the fermionic model 
with all the electrons frozen in half of the lattice
thus inhibiting charge transport, if $|V/t|>2$, and, notably, irrespectively
of whether the interaction is attractive or repulsive.

\acknowledgments

We thank M. Affronte, G. Falci, C. Hess and A. Parola 
for fruitful discussions.
T.P. acknowledges support by grants 
P1-0044 and J1-7347 of Slovenian Research Agency.

\end{document}